# Design of a Casimir-driven parametric amplifier

M. Imboden,[1,a)] J. Morrison,[2] D. K. Campbell[1,2,3] and D. J. Bishop[1,2,3]

[1]*Department of Electrical and Computer Engineering, Boston University, Boston, 02215, USA*

[2]*Department of Physics, Boston University, Boston, 02215, USA*

[3]*Division of Materials Science and Engineering, Boston University, Brookline, 02446, USA*

In this paper, we discuss a design for a MEMS parametric amplifier modulated by the Casimir force. We present the theory for such a device and show that it allows for the implementation of a very sensitive voltage measuring technique, where the amplitude of a high quality factor resonator includes a tenth power dependency on an applied DC voltage. This approach opens up a new and powerful measuring modality, applicable to other measurement types.

**I. INTRODUCTION: PARAMETRIC MEMS and the CASIMIR FORCE**

Parametric amplification is observed when a parameter of a resonator is modulated at twice the resonance frequency and in a controlled phase relation to the harmonic driving force. Given control of the modulation amplitude, frequency and phase, it is possible to amplify the mechanical oscillations until nonlinearities cap the otherwise diverging gain. This was demonstrated in a MEMS device in the work of Rugar and Grütter [1], who used a piezoelectrically driven cantilever beam that could be modulated electrically through capacitive coupling. They obtained a gain in the resonant ($\omega_0$) amplitude response of up to 100 by modulating a system parameter at $2\omega_0$ and at a phase difference between the drive and modulation of $\pi/2$. For a modulation with a zero phase difference, a factor of two deamplification was observed. The derivation of the parametric amplification followed work based on the normal-mode approach for electrical parametric oscillators by Louisell *et al.* [2].

Since then, mechanical parametric amplification has been observed in piezoelectric [3, 4], piezoresistive [5], and magnetomotive resonator systems [6], including high mechanical frequencies up to 130 MHz [7]. Optical pumping was used to mechanically amplify disc resonators [8], and capacitive-based modulations were used to parametrically amplify torsional resonators [9], MEMS [10], and low temperature NEMS devices [11]. Shu *et al.* [12] used the coupling of a doubly clamped beam to a qubit to parametrically modulate a NEMS resonator. The nonlinear behavior of the qubit resulted in more effective coupling when compared to the more conventional electrostatic methods. The mechanical gain lay typically in the range of 10-1000. In-depth calculations, including higher order terms, were considered in the work by Nasrolahzadeh *et al.* [13].

Here we discuss how using the Casimir force to parametrically modulate micro-electromechanical devices (MEMS) can lead to ultrasensitive sensing devices. We show that the strongly nonlinear forcing of the Casimir effect can be leveraged in a mechanical parametric amplifier to obtain extraordinarily high gain in a controlled and tunable way. We present a scheme for exploiting this coupling to measure small changes in a DC voltage signal by observing the dynamic response of a torsional oscillator. In our device, a voltage signal shifts the critical parameters of the parametric amplifier and thereby alters its gain. We exploit the strong dependence of the Casimir force on separation to obtain ultrahigh sensitivity through a $V^{10}$ dependence. Using electrostatic forces, our device presents a high impedance to the measured voltage.

---

[a)] Electronic mail: mimboden@bu.edu.

The DC voltage sensor is one of many possible configurations one could construct that use the Casimir-based parametric amplification to observe very small signals. Other potential applications include AC voltage sensors, AC or DC current sensors, and a detection scheme to measure displacements directly, all of which may be of interest in gravitational wave experiments. In all these cases, static, semi-static, or resonant displacements are monitored with a kHz voltage signal using a phase sensitive lock-in amplifier.

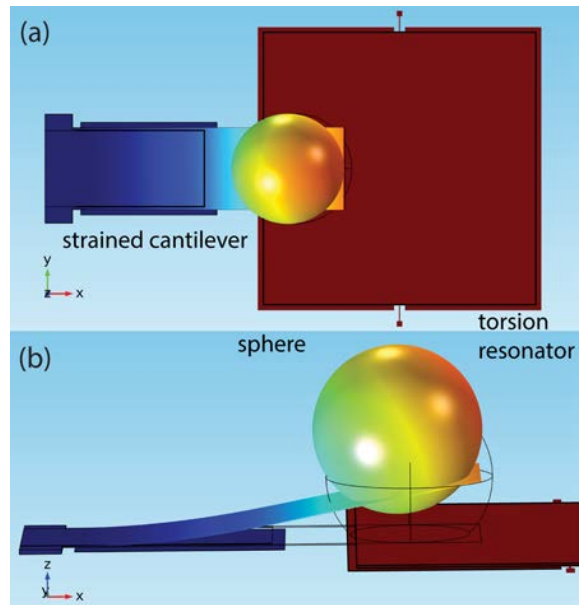

FIG. 1. Proposed MEMS device used to measure the static Casimir force. Top view (a) and side view (b) of strained cantilever, sphere and torsion resonator. The sphere and resonator are integrated MEMS devices fabricated on the same chip with micron alignment. The $z$ displacement is not to scale. The displacement of the cantilever and sphere is coded in color.

In essence, our technique allows one to detect a DC signal by measuring the amplitude of an AC signal at kHz frequencies. This method reduces sensitivity to electrical and mechanical noise which typically falls off as *1/f*. Following the introduction, we first derive the theory for Casimir-driven parametric amplification and then discuss a proposal for leveraging this amplification in a MEMS sensor.

The Casimir effect was first predicted in 1948 [14] as a result of deriving the Van der Waals force using retarded potentials. It predicts that conducting objects of the same potential feel an attractive force resulting from zero-point fluctuations of the electromagnetic field [14]. For small separations, this purely quantum mechanical force can be observed in classical macroscopic systems. According to quantum theory, the vacuum is never truly empty but is described as a froth of briefly appearing and self-annihilating zero-point fluctuations. Perfect (and also imperfect) conductors introduce a boundary condition that quantizes the possible fields that can exist between the two conductors. This results in a vacuum pressure between the conductors that is lower than the vacuum pressure of free space. As a result, the plates are pushed together. R. L. Jaffe [15] proposes a derivation of the Casimir force that does not include zero-point fluctuations. Though he does not contest that the zero-point fluctuations are probably real, he notes that: "Still, no known phenomenon, including the Casimir effect, demonstrates that zero-point energies are real." This issue does not affect any of the calculations regarding the Casimir effect in our manuscript. The Casimir effect was a contested theoretical idea until the development of experimental methods that could probe length scales well below 1 μm [16-19]. This intrinsically quantum mechanical force can now be used for practical applications, including as a non-contact displacement sensor [20]. Furthermore, Kenneth *et al.* [21] propose circumstances (essentially altering the boundary conditions) in which the Casimir force may be repulsive, further expanding the type of experiments and applications that can be conceived [22-24]. The Casimir force described here is assumed to be temperature independent. A significant amount of theoretical and, more recently, experimental effort [25] in describing the temperature-dependent Casimir force is worth noting. For this case, not only zero-point fluctuation but also thermal photons following Bose-Einstein statistics, are included. It is predicted that for the smallest separations, the temperature-independent Casimir force dominates, but due to the lower power scaling of the thermal bath, for larger separations, the temperature-dependent Casimir effect can dominate over the temperature-independent Casimir force [25]. Other experiments comparing measurements taken at 77 K and 300 K and smaller length scales have not been able to demonstrate thermal effects [26]. Klimchitskaya *et al.* [27] provide a comprehensive review of the theory and experiments of the Casimir force.

The approach discussed here is based on previous work on the static [18] and dynamic [19] detection of the Casimir force using a torsional resonator. The resonator is driven and detected electrostatically while brought within close proximity to a gold-coated sphere. The geometry considered is shown in Fig. 1. The derivation is analogous for the parallel plate configuration and is included for completion. The resulting Casimir force for the parallel plate ($F_{c\text{-}PP}$) [14] and the plate sphere ($F_{c\text{-}PS}$) [18] system can be expressed as:

$$F_{c-PP} = -\frac{\pi^2 \hbar c}{240} \frac{A}{z_0^4} \qquad (1a)$$

$$F_{c-PS} = -\frac{\pi^3 \hbar c}{360} \frac{R}{z_0^3}, \qquad (1b)$$

where $z_0$ is the plate-plate or plate-sphere separation, $A$ the area of the plates and $R$ the sphere radius. $\hbar$ is Plank's constant over $2\pi$ and $c$ is the speed of light in vacuum. Though the parallel plates exhibit a larger Casimir force, the alignment is considerably more difficult experimentally and hence generally the Casimir force is studied using the plate-sphere configuration. For non-perfect conductors, and taking surface roughness into consideration, would lead to some modifications to this expression, which, for simplicity we do not include in this initial study. These effects do need to be considered when analyzing an actual device. The theoretical work by Lamoreaux [28] was used to model the experiment [18] and agreed well with predictions.

This Casimir force can be detected through the deflection of the plate, whose angle is measured capacitively in a bridge-circuit detection scheme. In addition to the benefits of using a balanced-bridge detection technique, the resonant torsional degree of freedom of the device couples weakly to the mechanical noise from the environment, and hence allows for more accurate measurements. For a semi-static configuration, the deflection of the torsional structure can be detected capacitively as the metal sphere approaches the plate [18]. On resonance, the frequency of the torsional mode can be shifted and a nonlinear forcing term can be observed that is dependent on the sphere-plate separation [19]. Here, we propose to integrate the sphere-plate separation control on a single 2.5 mm × 2.5 mm die and to use MEMS-based technology to control and manipulate the Casimir-resonator interaction. An example of how this might be done is shown in Fig. 2. Modulating specific parameters will result in a Casimir-based parametric amplifier with strongly nonlinear coupling. The approach is similar to the MEMS-based Casimir sensor reported by Zou *et al.* [29]. The setup we propose differs in that it is still based on the well-studied sphere-plate setup. Furthermore, the surfaces used are gold or other metals of choice, instead of silicon, which approximate more closely ideal conductors.

The results reported in [29] demonstrate the strength of a MEMS-based approach and justify the methods proposed here. Recently, the Casimir force has been measured between a sphere and a vibrating membrane [30]. The surface potentials were mapped using a Kelvin probe. This allowed for high precision Casimir force measurements for separations ranging from 100 nm to 2 µm. In a theoretical work, Miri *et al.* [31] proposed using the lateral Casimir force to actuate a pinion. The nonlinear dynamics promises extremely sensitive devices, resulting from non-contact forces that preclude wear on the nanoscale devices. Our approach is a further application of the Casimir force, using the parametric amplification of the Casimir force to enable new sensing modalities with advantageous scaling behavior. This technique may help shed light on which models best describe the Casimir force. As the MEMS devices are compatible with cryogenic setups our device is suited for measuring the Casimir force over a wide range of temperatures.

TABLE I. Parametric modulation terms of electrostatic and Casimir forcing. $F_0 / \tau_0$ is the applied drive force/torque, and $\varphi$ is the phase difference of the forcing and modulation. $\frac{dC}{dz}$ and $\frac{d^2C}{dz^2}$ are the first and second derivatives of the capacitance along the axis of motion. $R$ is the sphere radius and $z$ the sphere plate separation; the subscript 0 indicates the static contribution and the subscript $p$ the parametric modulation. $\hbar$ is Planck's constant over $2\pi$ and $c$ the speed of light in vacuum.

| Term | Electrostatic [1] | Casimir (Sphere-linear Plate) | Casimir (Sphere-torsion Plate) | Casimir (Parallel-Plate) |
|---|---|---|---|---|
| Harmonic forcing/torque: $F_{dr}(t)/\tau_{dr}(t)$ | $F_0 \cos(\omega t)$ (Piezoelectric) | $F_0 \cos(\omega t)$ (Electrostatic) | $\tau_0 \cos(\omega t)$ (Electrostatic) | $F_0 \cos(\omega t)$ (Electrostatic) |
| Parametric force/torque: $F_p(t)/\tau_p(t)$ | $\frac{1}{2}\frac{dC}{dz}V(t)^2$ | $-\frac{\pi^3 \hbar c}{360}\frac{R}{z(t)^3}$ | $-\frac{\pi^3 \hbar c}{360}\frac{RL_{Sph}}{z(t)^3}$ | $-\frac{\pi^2 \hbar c}{240}\frac{A}{z(t)^4}$ |
| Modulated spring constant: $k_p(z,t)/\kappa_p(z,t)$ | $\frac{1}{2}\frac{d^2C}{dz^2}V(t)^2$ | $\frac{\pi^3 \hbar c}{120}\frac{R}{z(t)^4}$ | $\frac{\pi^3 \hbar c}{120}\frac{RL_{Sph}^2}{z(t)^4}$ | $\frac{\pi^2 \hbar c}{60}\frac{A}{z(t)^5}$ |
| Modulation parameter | $V(t) = V_0 + V_P \cos(2\omega t + \varphi)$ | $z(t) = z_0 + z_P \cos(2\omega t + \varphi)$ | $z(t) = z_0 + z_P \cos(2\omega t + \varphi)$ | $z(t) = z_0 + z_P \cos(2\omega t + \varphi)$ |
| Modulation amplitude $\Delta k/\Delta \kappa$ [a] | $\frac{d^2C}{dz^2}V_0 V_P$ | $-\frac{\pi^3 \hbar c}{30}\frac{R z_p}{z_0^5}$ | $-\frac{\pi^3 \hbar c}{30}\frac{RL_{Sph}^2 z_p}{z_0^5}$ | $-\frac{\pi^2 \hbar c}{60}\frac{A z_p}{z_0^6}$ |
| Static term | $\frac{1}{2}\frac{d^2C}{dz^2}V_0^2$ | $\frac{\pi^3 \hbar c}{120}\frac{R}{z_0^4}$ | $\frac{\pi^3 \hbar c}{120}\frac{RL_{Sph}^2}{z_0^4}$ | $\frac{\pi^2 \hbar c}{60}\frac{A}{z_0^5}$ |
| Max parametric gain | $\frac{1}{1-\frac{Q}{2k_0}\frac{d^2C}{dz^2}V_0 V_P}$ | $\frac{1}{1-\frac{Q\pi^3 \hbar c}{60 k_0}\frac{R z_p}{z_0^5}}$ | $\frac{1}{1-\frac{Q\pi^3 \hbar c L_{Sph}^2}{60 \kappa_0}\frac{R z_p}{z_0^5}}$ | $\frac{1}{1-\frac{Q\pi^2 \hbar c}{120 k_0}\frac{A z_p}{z_0^6}}$ |

[a] Terms of $k_p(z,t)$ and $\kappa_p(z,t)$ proportional to $\cos(2\omega t)$, DC terms causing static offset, and higher harmonics are suppressed for the case where $V_0 >> V_P$ and $z_0 >> z_P$.

## II. CASIMIR-DRIVEN PARAMETRIC AMPLIFIER

Conceptually, our procedure is modeled on the experiment by Rugar and Grütter [1], in which they use electrostatic coupling to drive the MEMS parametric amplifier. Here we show how analogously to the electrostatic force, the Casimir force can be used parametrically to amplify the resonant response. The unusual scaling of the Casimir force enables unique sensitivity characteristics. We propose to use this effect to monitor changes in a DC voltage. We predict how a DC voltage signal can be measured by coupling it to the resonant response of a mechanical torsion oscillator. Whereas in an electrostatic drive and detection scheme the resonant amplitude scales as $V^2$, coupling the DC voltage to the resonator through a parametrically modulated Casimir force results in a $V^{10}$ amplitude dependence. The ultimate sensitivity of the device will depend on the ability to tune the parameters close to where the gain diverges in the linear approximation (although eventually nonlinearities of some sort will saturate the gain at a finite magnitude). A MEMS-based device can be tuned to maximize this effect. Multiple parameters, such as the DC and AC voltages of the resonator and cantilever, can be adjusted and tuned so that the parametric amplification occurs at the ideal place in parameter space. To demonstrate how this may work, we derive here the Casimir parametric amplification in parallel with the well-established electrostatic modulation described by Rugar and Grütter [1]. While not new, this derivation is essential to understanding the potential advantages of the Casimir approach.

By including a term for the parametric modulation of the spring constant, the equation of motion of a damped driven harmonic oscillator becomes:

$$m\ddot{z}_a + \frac{m\omega_0}{Q}\dot{z}_a + [k_0 + k_p(z,t)]z_a = F_{dr}(z,t), \qquad (2)$$

where $z_a$ is the amplitude of the resonator, $m$ the mass, $Q$ the quality factor (for these types of MEMS $Q$ is typically large and a resonance is sustained by the harmonic drive force), $F_{dr} = F_0\cos(\omega t)$ the drive force close to $\omega_0$, $\omega_0$ the resonance frequency and $k_0 = m\omega_0^2$ the unperturbed spring constant. (A description of a capacitively driven, damped torsional oscillator, including nonlinear terms is given by [32] and included in appendix C. For simplicity all expressions derived here are for the linear spring model. As the actuation amplitude is small, the conversion to torque and angles from forces and amplitudes is trivial). For harmonic electrostatic actuation $F_0 = \frac{dC}{dz}V_{DC}V_{AC}$, where $\frac{dC}{dz}$ is the derivative of the capacitance, $V_{DC}$ the DC bias and $V_{AC}$ the harmonic potential. The perturbation can be expressed as the derivative of an external forcing term:

$$k_p(z,t) = \frac{dF_P(z,t)}{dz} \qquad (3)$$

where $F_P$ is the parametric perturbation force defined in Table I, which considers all the terms as described by [1] for the electrostatic modulation, as well as for the proposed Casimir modulation.

The unperturbed sphere-resonator separation, $z_0$, is not truly static, but oscillates with the amplitude of the torsional resonator. Hence $z_0 \rightarrow z_0 + z_a\cos(\omega_0 t)$. The parametric perturbation amplitude, $z_P$, is on the same order of magnitude as $z_a$. The contribution to the parametric modulation due to the motion of the plate is suppressed by order $\left(\frac{z_a}{z_0}\right)^2$ compared to the intended modulation. The harmonic forcing is also shifted by a term proportional to $\frac{z_a}{z_0}$. Again, in the limit where $z_0 >> z_a \approx z_P$, this term can be neglected.

The resulting gain due to parametric amplification can be expressed as [1]:

$$G = \left[\frac{\cos\varphi^2}{\left(1+\frac{Q\Delta k}{2k_0}\right)^2} + \frac{\sin\varphi^2}{\left(1-\frac{Q\Delta k}{2k_0}\right)^2}\right]^{\frac{1}{2}}, \qquad (4)$$

and is defined as the ratio of the on-resonant amplitude response with and without parametric modulation ($G = \frac{|z|_{mod\,on}}{|z|_{mod\,off}}$). Note that for the Casimir parametric amplifier, $\Delta k = -\frac{\pi^3 \hbar c}{30}\frac{Rz_p}{z_0^5} < 0$ (plate-sphere) and $\Delta k = -\frac{\pi^2 \hbar c}{60}\frac{A\,z_p}{z_0^6}$ (parallel plate). This means that a maximum gain is observed for $\varphi = 0$, and a maximum de-amplification is observed for $\varphi = \pi/2$, the opposite phases compared to the electrostatic parametric modulation, where $\Delta k = \frac{d^2C}{dz^2}V_0V_P > 0$.

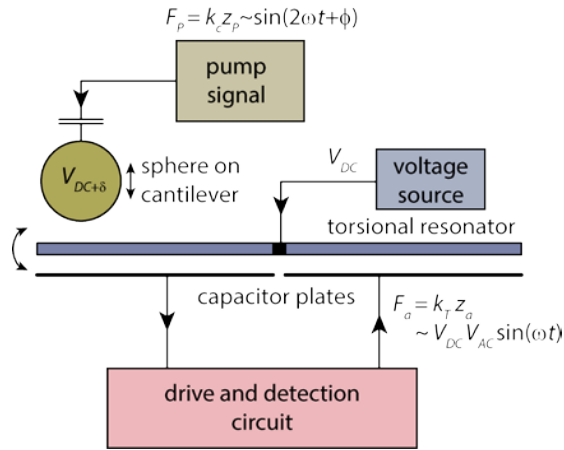

FIG. 2. Block diagram of the Casimir parametric amplifier. The parametric modulation occurs at twice the frequency of the harmonic drive. $\delta$ corresponds to the offset voltage required to cancel the electrostatic forces caused by potential islands in the polycrystalline metals.

Electrostatically, the parametric modulation amplitude can be increased by increasing the DC voltage $V_0$ or the AC voltage $V_p$ and each scales linearly with $\Delta k$. Using the Casimir modulation, $k_P(z,t)$ scales with the displacement as $\sim (z_0 + z_P)^{-4} \approx z_0^{-4}\left(1 - 4\frac{z_P}{z_0} + O\left(\left(\frac{z_P}{z_0}\right)^2\right)\right)$. This is the term that gives the Casimir parametric amplifier extraordinary displacement sensitivity. For a displacement modulation at a frequency of $2\omega$, the gain will depend on the second term in the above Taylor series expansion. This results in a power dependence of $z_0^{-5}$. Table AI in Appendix A gives numerical examples for the electrostatic as well as predicted values for the Casimir parametric amplification. It is estimated that $\frac{Q}{2k_0}$ will range between $10^3$ and $10^4$ m/N. For a divergent gain, $\Delta k$ must then be tunable from 0 to $10^{-4}$ N/m. Given typical sphere and resonator dimensions (summarized in Tables AII and AIII in Appendix A), the required values of $\Delta k$ correspond to a modulation amplitude $z_p$ of 3 nm, given a sphere-plate separation $z_0$ of 100 nm. The contact potential can be nulled by finding the minimizing potential to reduce the potential difference between the gold sphere and torsional resonator. The remaining electrostatic force due to the patch potentials, characterized by the root-mean-square potential $V_{rms}$, must be accounted for [33]. These are impossible to predict theoretically as they are sensitive to the fabrication process and must be measured experimentally. Ideally the residual electrostatic forces are small and the electrostatic parametric modulation $\Delta k$ vanishes, so that the parametric amplification is a consequence of the perturbation due to the Casimir force alone. In real experiments a contribution of $\Delta k_{el} = \frac{d^2C}{dz^2}V_{rms}^2$ will be added, and if $V_{rms}$ is large compared to $\sqrt{\frac{\pi^3 \hbar c}{30}\frac{Rz_p}{z_0^5}/\frac{d^2C}{dz^2}}$ the electrostatic parametric gain will overpower the contribution due to the Casimir force. The difference in scaling of the electrostatic and Casimir parametric modulation can help disentangle the respective contributions. Whereas patch potentials, surface roughness and the finite conductivity of gold adjusts the scaling of the forces between the sphere and the plate [34], the actual achievable gain is of course limited by nonlinearities in the system. These could include nonlinearities in the spring constant described by the Duffing equation, where typically geometric and external potential effects dominate over intrinsic material nonlinearities [35]. Nonlinear dissipation, observed in nano-electromechanical systems [36], would also limit the achievable gain. This mechanism has been identified as the probable limiting cause in parametrically amplified mechanical resonators [12, 37]. Attempting to model these many different nonlinear effects before the actual devices are constructed is of limited value, but understanding them in the actual devices will be critical to optimizing the designs.

We predict a Casimir pull-in effect analogous to the snap-down observed in capacitive devices. As the attractive Casimir force increases nonlinearly with diminishing separation, it will eventually grow faster than the linear restoring force arising from the spring constant given by Hooke's Law. The consequence of this is a finite stable minimum separation of the ball-resonator system. This will set a natural lower bound on the separation $z_0$, which consequently sets the maximum sensitivity predicted for this type of amplifier. The pull-in characteristics due to electrostatic and Casimir forces for cantilever and doubly clamped beams have been calculated by [38] resulting in detachment lengths for MEMS/NEMS (the minimum allowed length of a cantilever or doubly clamped beam below which the Casimir force inevitably leads to collapse). The effects of surface roughness on the pull-in are described by [39]. It has been demonstrated that patterned features smaller than the plasma wavelength strongly suppress the Casmir forces [40] and would hence also suppress pull-in effects. Here results of the lumped model analysis are presented with the assumption that all electrostatic forces have been nulled and the pull-in is entirely a result of the Casimir force. In future experiments the residual electrostatic forces [41], surface roughness and finite conductivity contributions [42] will need to be included in the model.

The closest stable approach before pull-in occurs when the generalized spring constant vanishes:

$$k_{sys} = -\frac{\partial F_{tot}}{\partial z} = \frac{\pi^3 \hbar c}{120} \frac{R}{(z_0-z)^4} - k_0 = 0. \tag{5}$$

The spring restoring force is balanced by the attractive Casimir force ($F_{tot} = 0 = \frac{\pi^3 \hbar c}{360} \frac{R}{(z_0-z)^3} + k_0 z$). Hence, using $k_0 = \frac{\pi^3 \hbar c}{360} \frac{zR}{(z_0-z)^3}$, the solution to equation (5) is $z = \frac{z_0}{4}$. Where the electrostatic force experiences the well-known $^1/_3$ pull-in effect, the Casimir attractive force results in a ¼ pull-in. Electrostatically, the pull-in can be observed by increasing the bias voltage, resulting in the maximum-allowed voltage of $V_P = \sqrt{\frac{8}{27} \frac{k_0 z_0^3}{\varepsilon_0 A}}$. Here, there is no such tunable parameter, but the pull-in is of critical importance, as it defines the minimum separation a conducting plate-sphere setup can have before the Casimir force pulls the two together. Substituting $z = \frac{z_0}{4}$ in eqn. (5) and solving for $z_0$, one obtains:

$$z_{0C} = \sqrt[4]{\frac{32\pi^3 \hbar c}{1215} \frac{R}{k_0}} = \gamma_{CP} \sqrt[4]{\frac{R}{k_0}}, \tag{6}$$

where $\gamma_{CP} = \sqrt[4]{\frac{32\pi^3 \hbar c}{1215}} \cong 3.956 \times 10^{-7}$ N$^{\frac{1}{4}}$m$^{\frac{1}{2}}$ is the pull-in constant for the Casimir-spring sphere-plate system. Using typical numbers for the setup proposed here ($R$ = 100 μm, $k_0$ =0.2 N/m) results in a minimum separation of the sphere plate of $z_{0C}$ = 60 nm. The derivations for a linear spring parallel plate setup and for a plate-sphere torsional structure are included in Appendix D.

The minimum theoretical value for $z_0$ is dictated by the Casimir pull-in separation described above. For large gain, the amplitude $z_a$ must be limited so that the condition $z_a << z_0$ still holds (otherwise higher order terms in the derivation of the mechanical response need to be considered). Naturally $z_a$ must always be greater than the mechanical thermal noise amplitude $z_{th}$, resulting in a condition for the minimum possible $z_a$ and corresponding maximum allowed gain. As discussed in Appendix E, the thermal-mechanical on-resonance angular amplitude for the torsion resonator with a rectangular rod considered here scales as $\sqrt{T}$. Multiplying the spectral noise density by the bandwidth of the resonator results in a room temperature thermal amplitude of $\Delta z_{th} \cong 38$ pm. At $T$ = 4.2 K the thermal amplitude is $\Delta z_{th} \cong 4$ pm. Interesting thermal effects, including self-oscillations, might be observable, but initial experiments would be conducted in the more conservative parameter space constrained thermally by the minimum amplitude as discussed in Appendix E.

## III. MEMS-BASED CASIMIR VOLTAGE SENSING

We propose to build an integrated device that uses a strained cantilever beam to suspend a gold sphere above the MEMS torsion resonator. This self-aligned setup would allow for precise tuning of all relevant parameters, resulting in a compact, mechanically stable device that is suitably engineered to study Casimir forces and the Casimir-based parametric amplification described above. One of multiple sensing applications includes high-impedance DC voltage measurements. Alternatively, low-impedance current sensing, thermal sensing, and displacement sensing could also be envisioned. In all cases, the desired signal is converted into a small displacement of the cantilever. High signal sensitivity is obtained by tuning the device to a specific parameter space, close to where in the absence of nonlinear effects the parametric gain diverges. In essence, the mechanical gain observed can be perceived as a mechanical amplification of the Casimir force, i.e., the gain manifests itself as an increase in the strength of the Casimir force.

Fig. 3 schematically depicts the integrated MEMS system, including actuation, sensing and modulation electrodes. A DC bias voltage nudges the sphere towards the plate. In conjunction with the parametric amplification setup, the amplitude of the resonance response of the oscillator will increase as the gain increases. The following derivation demonstrates the high sensitivity of the device, as the parametric gain in this setup is sensitive to the 10$^{th}$ power of the applied voltage. How such a device may be fabricated and calibrated is outlined in Appendix B. A parallel plate setup would result in a 12$^{th}$ power sensitivity to an applied voltage; however due to the high level of control needed for the setup to work a plate-sphere configuration is probably the more robust approach.

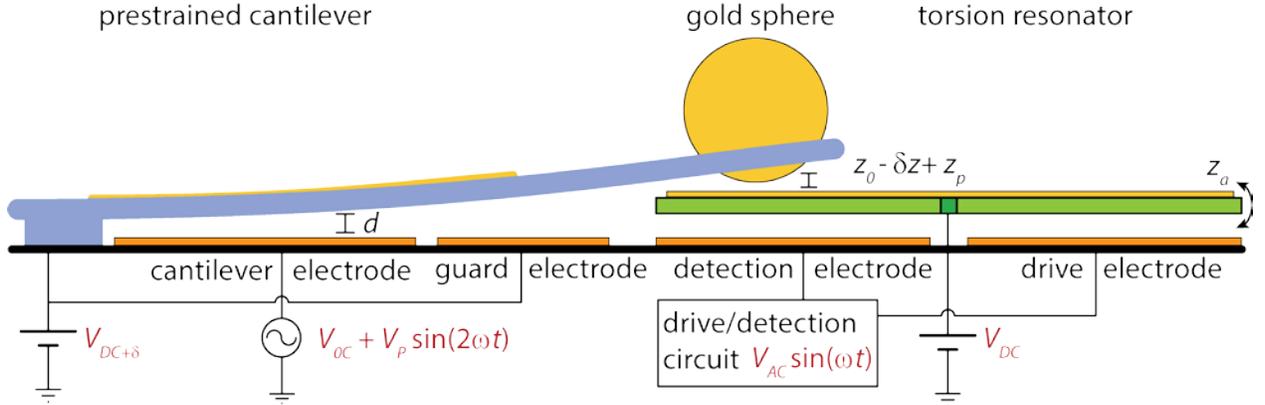

FIG. 3. Torsion resonator with gold sphere suspended above by a strained cantilever. The height of the sphere can by adjusted by applying a voltage to an electrode beneath the cantilever. $\delta$ corresponds to the offset voltage required to cancel the electrostatic forces caused by potential islands in the polycrystalline metals.

The cantilever deflection for a uniform load per unit area ($f_A$) is given by:

$$\delta z = \frac{f_A L_c^4}{8 E I_c} = \frac{f_A L_c}{k_c}, \tag{7}$$

where $E = 150$ GPa is the Young's modulus of polysilicon, $I_c = \frac{w_c t_c^3}{12}$ the second moment of inertia of a rectangular beam, and $k_c = \frac{8 E I_c}{L_c^3}$ the spring constant of the cantilever. $L_c$, $w_c$, and $t_c$ parameterize the length, width, and thickness of the cantilever suspending the gold sphere. The capacitive forcing per unit area for small displacements ($\delta << d$) of two parallel plates is given by:

$$f \approx -\frac{1}{2} \varepsilon_0 \frac{V_{0C}^2}{d^2}. \tag{8}$$

It is assumed that the deflection caused by the voltage $V_{0c}$ applied to an electrode beneath the cantilever does not significantly change the capacitor gap size $d$. The cantilever displacement as a function of applied voltage results in:

$$\delta z = -\frac{3\varepsilon_0 L_c^4}{4E t_c^3}\frac{V_{0c}^2}{d^2} = -\frac{\varepsilon_0 w_c L_c}{2k_c}\frac{V_{0c}^2}{d^2}$$

$$= -\alpha_c V_{0c}^2, \tag{9}$$

where $\alpha_c$ is the electromechanical coupling strength of the cantilever. At zero voltage, the sphere is placed at a distance of $z_0$ above the resonator. In this configuration, the maximum gain is given by (eqn. (4), using $\varphi = 0$):

$$G = \frac{1}{\left(1 - \frac{Q\pi^3 \hbar c R z_p}{60 k_0 z_0^5}\right)}. \tag{10}$$

The $z_p$ modulation can be added through the resonator bias or on the cantilever as depicted in Fig. 2. The $z_0$ displacement becomes a function of the cantilever bias voltage $z_0 \to z_0 + \delta z(V_{0c})$. Hence, the gain becomes:

$$G = \frac{1}{\left(1 - \frac{Q\pi^3 \hbar c R z_p}{60 k_0 (z_0 - \alpha_c V_{0c}^2)^5}\right)}, \tag{11}$$

where $Q$ and $k_0$ are the quality factor and spring constant of the resonator respectively.

Maximum sensitivity of the DC voltage $V_{0c}$ is obtained for greatest changes in gain. To find this region of parameter space, we differentiate the gain with respect to the applied voltage $V_{0c}$ and linearize to determine the leading term:

$$\frac{\partial G}{\partial V} = \frac{5\Delta k z_0^5 Q \alpha_c V}{k_0 (z_0 - \alpha_c V^2)^6 \left(1 - \frac{1}{2k_0}\frac{Q|\Delta k|z_0^5}{(z_0 - \alpha_c V_{0c}^2)^5}\right)^2} \approx \frac{10 Q|\Delta k|\varepsilon_0}{d^2 k_c z_0 (Q|\Delta k| - 2k_0)^2} V_{0c} + O(V_{0c}^3). \tag{12}$$

Specifically, for the Casimir modulated spring constant, we find:

$$\frac{\partial G}{\partial V} = \frac{\pi^3 \hbar c Q R z_p \alpha_c V}{6k_0 (z_0 - \alpha_c V^2)^6 \left(1 - \frac{\pi^3}{60 k_0}\frac{\hbar c Q R z_p}{(z_0 - \alpha_c V_{0c}^2)^5}\right)^2} \approx \frac{600\, \pi^3 \hbar c Q R z_p z_0^4 k_0 \alpha_c}{(\pi^3 \hbar c Q R z_p - 60 k_0 z_0^5)^2} V_{0c} + O(V_{0c}^3), \tag{13}$$

The sensitivity, even to first order, diverges as $60 k_0 z_0^5 \to \pi^3 \hbar c Q R z_p$, resulting in the amplitude relation for maximum sensitivity:

$$\frac{z_p}{z_0^5} = \frac{60 k_0}{\pi^3 \hbar c Q R}. \tag{14}$$

This implies that any arbitrary sensitivity can be obtained by tuning the static and parametrically modulated amplitudes. Table AI includes numerical values that are experimentally feasible for the parameters considered here, noting that $z_0 > z_{0C}$ and $z_a > z_{th}$.

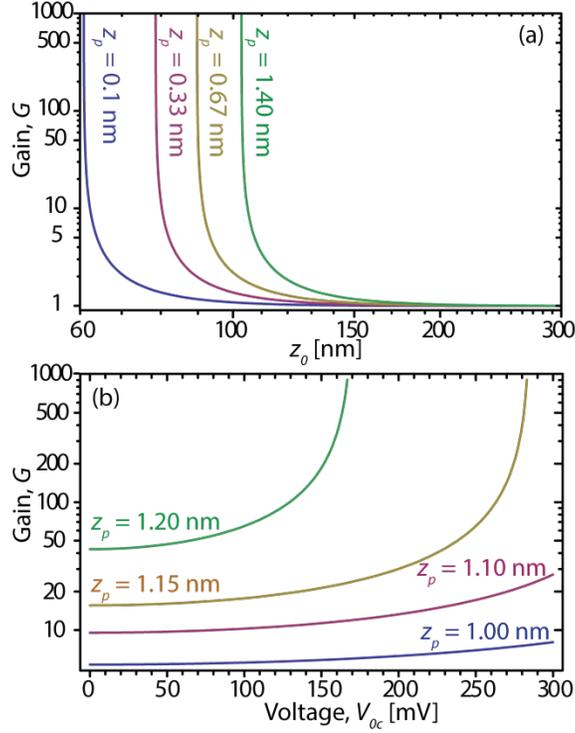

FIG. 4. Gain dependency on sphere-plate separation (a) and the voltage (b) for changing modulation amplitude $z_P$. All other parameter values are given in Tables AII and AIII of the Appendix.

Fig. 4 depicts the gain dependency observed on resonance for sweeping DC bias voltages. Three different modulation amplitudes are considered, $z_0$, $Q$, $k_0$, $R$ and the cantilever parameters are the same as in Table AII. As becomes evident, with sufficient control and tuning of the setup, the gain and hence the sensitivity can always approach the point of divergence. The ultimate sensitivity that can be obtained will depend on various nonlinear effects and must be determined empirically. The parametric gain changes the angular amplitude of the torsional resonator. The voltage sensitivity will depend on both the gain resulting from the parametric modulation and the accuracy to which the amplitude of the resonator can be measured.

## IV. DISCUSSION AND CONCLUSIONS

We have demonstrated a model for a Casimir-driven parametric amplifier. We illustrate how the nonlinear modulation resulting from the Casimir force can be leveraged to parametrically amplify a harmonic torsional resonator. We analyze specifically a setup for DC voltage detection, which scales as $V_{0c}^{10}$. We predict high sensitivity within a tunable range. In a fully linear system, the gain diverges; in real systems, nonlinear effects including dissipation or anharmonicity in the spring constant are expected to limit the maximum gain and hence the sensitivity. As the gain diverges in principle any sensitivity could be achieved. Of course in practice this cannot be true. Other published parametric gains for MEMS and NEMS devices typically range from 10-1000. In some cases nonlinearities and self-generation [8, 11] and nonlinear dissipation [12, 43] may be the limiting factor, in others the stability and tunability to that the threshold, as the gain also amplifies the noise of the system [44]. Given the crystalline nature of both the gold on the torsional resonator and the gold sphere, patch potentials and finite conductivity effects will likely further limit gain and sensitivity and alter the $V_{0c}^{10}$ dependence. The patch potentials can be measured and nulled by minimizing the electrostatic force acting on the resonator [18, 29]; calculations for plate-sphere geometries are given by [45]. In practice this may lead to a separation-dependent correction potential that must be added dynamically to the gold sphere via the cantilever. Though the force due to patch potentials varies with separation [46], the scaling is very different from the Casimir force and hence it may be possible to decouple the electrostatic perturbations from the Casimir parametric amplification. Experiments are clearly best suited to determine the actual gain and sensitivity of such a setup.

The method proposed is applicable to other measurement configurations, including AC voltage measurement, low-impedance current measurements, temperature sensing and displacement sensing. Apart from sensing applications, the parametric gain acts as a gain in the coupling strength of the Casimir force, making the device an interesting platform for Casimir measurements. An integrated MEMS device may enable significant improvements in the sensitivity of Casimir experiments. This could shed light on experiments studying material, surface topology [47, 48] and metamaterials [49, 50], an understanding of which is crucial for stabilizing nano-structures subjected to the Casimir force. Ultimately, using parametrically driven MEMS may even enable high precision measurements of non-Newtonian forces acting on short length scales [51].

## ACKNOWLEDGMENTS

This research is funded in part by Boston University.

## APPENDIX

## APPENDIX A: Typical MEMS Parameter Values

TABLE AI. Typical numbers used in electrostatic parametric amplification and predicted magnitudes for the proposed Casimir parametric amplification. The factor $\frac{Q}{2k_0}$ can vary over a considerable range by adjusting the spring constant $k_0$ with geometry and $Q$ by tuning the bias voltage, temperature or medium. For each case the modulation amplitude range is given for the full range of possible parametric amplification.

| Term | Electrostatic [1] | Casimir (sphere-linear plate) |
|---|---|---|
| $\frac{Q}{2k_0}$ | ~5000 | 1000 - 10000 |
| $\lvert \Delta k \rvert$ | $C''V_0V_P = 0 - \frac{2k_0}{Q}$ $= 0 - 0.0002$ $V_0 = 10$ V $V_P = 0\text{-}2.5$ V | $\frac{\pi^3 \hbar c}{30} \frac{R z_p}{z_0^5} = 0 - \frac{2k_0}{Q}$ $\approx 0 - (0.0001, 0.001)$ for $z_0 = 100$ nm and $z_p = 0\text{-}(0.3, 3)$ nm for $z_0 = 50$ nm and $z_p = 0\text{-}(0.01, 0.1)$ nm |

Table AII. Mechanical Parameters.

| Parameter | Notation | Value |
|---|---|---|
| Quality Factor | $Q$ | 1000 |
| Width of spring | $ws$ | 3 μm |
| Length of spring | $L_S$ | 40 μm |
| Thickness of spring | $t_s$ | 2 μm |
| Shear modulus of silicon | $G_{si}$ | 69 GPa |
| Young's Modulus of Silicon | $E$ | 150 GPa |
| Length of plate | $L_P$ | 500 μm |
| Position of sphere from center of plate | $L_{Sph}$ | 175 μm |
| Effective spring constant of plate | $k_0$ | 0.2 N/m |
| Width of cantilever | $w_c$ | 150 μm |
| Length of cantilever | $L_c$ | 500 μm |
| Spring constant of cantilever | $k_c$ | 1.1 N/m |
| Cantilever electrode separation | $d$ | 2 μm |
| Numeric constant of the torsion spring | $\beta$ | 2.433 |

Table AIII. Expected numerical values for a MEMS parametric amplifier.

| Parameter | Value |
|---|---|
| $R$ | 100 μm |
| $z_0$ | 100 nm |
| $z_P$ | 1.2 nm |
| $G(V_{0c} = 0)$ | 43 |
| $G(V_{0c} = 100$ mV$)$ | 65 |

## APPENDIX B: MEMS Fabrication and Calibration

Previous experiments have demonstrated that MEMS manufactured using standard optical lithography are sufficiently sensitive and stable to detect the Casimir force [18, 19]. The multi-user project run PolyMUMPs [52] enables reproducible and cost effective fabrication of the resonator-cantilever system. The strain in the cantilever is provided by a gold layer. The strain of both the poly-silicon and gold can be measured independently. Further annealing the devices can tune the height to a desired position.

The gold sphere is suspended over the torsional resonator in a hole fabricated close to the end of the cantilever. The sphere and hole radii determine the extrusion of the sphere beneath the cantilever as visible in Figs. 1(b) and 3, given by

$$h = R\left(1 - \sqrt{1 - \left(\frac{r_h}{R}\right)^2}\right) - t_c, \tag{B1}$$

where $R$ is the sphere radius, $r_h$ is the radius of the hole in the cantilever and $t_c$ is the cantilever thickness. The final sphere-plate separation $z_0$ is a combination of the extrusion height $h$, resonator height above the substrate (~5 μm in our case) and the height of the strained cantilever, given by

$$h_c = \frac{L_c^2}{2\,r_c}, \tag{B2}$$

assuming a large radius of curvature $r_c$. $L_c$ is the length of the cantilever. The radius of curvature is given by

$$\frac{1}{r_c} = \frac{6\varepsilon(T)w_m w_s E_m E_s t_m t_s (t_m + t_s)}{(w_m E_m t_m)^2 + (w_s E_s t_s)^2 + 2w_m w_s E_m E_s t_m t_s (2t_m^2 + 3t_m t_s + 2t_s^2)}. \tag{B3}$$

$w$, $t$, and $E$ denote the width, thickness and Young's modulus for silicon ($s$) and gold ($m$) respectively. $\varepsilon(T)$ is the temperature dependent strain. The strain results in part due to the different thermal expansion of polysilicon and gold. Hence a significant thermal dependence is expected. For stable operation the temperature of the die will need to be controlled. Using on-chip heaters and thermometers the device temperature can be stabilized to within 1 mK [53].

The placement of the gold sphere is achieved post release. This can be done using piezo-actuated micromanipulators. Depending on the required precision, micro-assembly can be completed in an SEM or FIB, using electron beam cured glue or FIB deposited platinum.

For the derivation given previously, the gain is determined for the case where the cantilever and actuating electrode can be treated as an idealized parallel plate capacitor. For a large radius of curvature and a well-placed sphere this approximation is valid. As the radius on curvature decreases and the required height tuning increases, a more involved calibration needs to be performed. An optical interferometer can be used to map out the amplitude as a function of voltage, giving a quantitative result for the electromechanical transfer function

$$d = \frac{1}{2k} \frac{dC}{dz} V_{DC}^2, \tag{B4}$$

where $d$ is the displacement, $k$ the spring constant and $V_{DC}$ the actuation voltage. $\frac{dC}{dz}$ is the derivative of the capacitance along the axis of motion which itself becomes a function of the applied voltage

The resonance frequency and displacement of both the torsional resonator and the cantilever can be measured using standard methods [19].

## APPENDIX C: The damped driven torsional oscillator modulated by a Casimir torque

The main text describes parametric amplification due to the Casimir Force in terms of a linear spring system. Experimentally there are significant advantages to use a torsional resonator. These include improved decoupling form mechanical noise and high sensitivity bridge detection schemes. The Casimir force acting between a plate and a sphere is modeled using the proximity force approximation [54], where the sphere is approximated as a sum of infinitesimally small parallel plates at varying vertical distances from the torsional plate. The approximation is valid for small plate-sphere separations compared to the radius if the sphere ($z/R<<1$).

The differential equation for a parametrically modulated damped driven torsion resonator is expressed as (equivalent to equation (2) for the linear system):

$$I\ddot{\theta} + \frac{I\omega_0}{Q}\dot{\theta} + [\kappa_0 + \kappa_p(\theta, t)]\theta = \tau_{dr}(\theta, t), \tag{C1}$$

where $I$ is the moment of inertia, $Q$ the quality factor, $\omega_0$ the angular resonance frequency, $\kappa_0$ the torsion spring constant and $\kappa_p$ the parametrically modulated torsion spring constant. $\theta$ is the deflection angle and $\tau_{dr}$ the harmonic drive, which in this case is electrostatic.

In order to derive the Casimir torque between the sphere-torsional plate setup one starts with the energy per unit area of a parallel plate system [14], given by:

$$\varepsilon(z) = -\frac{\pi^2 \hbar c}{720 \, z^3}. \tag{C2}$$

$z$, as above, is the parallel plate separation. The total potential energy can be defined for the plate-sphere system as a sum of infinitesimal parallel plates. Consequently, the total Casimir energy is the integral over the hemisphere facing the plate:

$$U = \int_0^{2\pi} R \, d\phi_s \int_0^{\frac{\pi}{2}} \varepsilon(z_0 + R(1 - \cos\theta_s)) R \, \sin\theta_s d\theta_s. \quad (C3)$$

$z_0$ is the point of closest approach, $R$ is the sphere radius and $\theta_s$ and $\phi_s$ parameterize the hemisphere. The torque on the plate due to the Casimir force can be calculated as the force between infinitesimal parallel plates, given by the derivative of the energy $\frac{dU}{dz}$, multiplied by the moment arm, $L_{Sph} - R \, \sin(\phi_s)$.

$$\tau_{C-PS} = \int_0^{2\pi} \int_0^{\frac{\pi}{2}} R^2 \, \sin\theta_s \, (L_{Sph} - R \, \sin(\phi_s)) \frac{d\varepsilon(z)}{dz} d\phi_s d\theta_s =$$
$$\frac{\pi^2 \hbar c}{120} R^2 \int_0^{2\pi} \int_0^{\frac{\pi}{2}} \frac{\sin\theta_s \, (L_{Sph} - R \, \sin(\phi_s))}{(z_0 + R(1-\cos(\theta_s)))^4} d\phi_s d\theta_s, \quad (C4)$$

where $z = z_0 + R(1 - \cos(\theta_s))$ was used. $L_{Sph}$ is the distance from the point of closest approach to the axis of rotation of the plate. The result of the above integral yields a straightforward expression:

$$\tau_{C-PS}(z_0) = -\frac{\pi^3 \hbar c \, R^2 L_{Sph}}{360 \, z_0^3} \frac{(R^2 + 3Rz_0 + 3z_0^2)}{(R+z_0)^3} \approx -\frac{\pi^3 \hbar c \, RL_{Sph}}{360 \, z_0^3} = F_{C-PS} L_{Sph}. \quad (C5)$$

where it is again assumed that ($z/R \ll 1$).

The parametric modulation is again expressed as the derivative of the external torque due to the Casimir effect. (Equivalent to equation (3) for the linear spring system.) The modulated torsion constant results from the derivative of the Casimir torque:

$$\kappa_P = \frac{d\tau_P(z,t)}{dz} \frac{dz}{d\theta} \approx \frac{d}{dz}\left(\frac{-\pi^3 \hbar c R L_{Sph}}{360 \, z^3}\right) L_{Sph} = \frac{\pi^3 \hbar c \, R L_{Sph}^2}{120 \, z^4}. \quad (C6)$$

The presence of the sphere introduces a static shift in $\theta$ in addition to the modulation resulting from the modulation of z: $z(t) = z_0 + z_P \cos(2\omega t + \varphi)$. The modulation amplitude (occurring at a frequency of $2\omega$) becomes

$$\Delta\kappa_P = -\frac{\pi^3 \hbar c \, R L_{Sph}^2}{30} \frac{z_P}{z_0^5} \quad (C7)$$

As the modulation is the result of the linear amplitude $z$. Given that for all amplitudes in $z$ considered here $\theta \approx z/L_{Sph}$, the mapping between the torsional and linear spring setup is straight-forward. Considering a linear force acting at a distance $L_{Sph}$ from the torsion spring, the torsion resonator responds with an effective linear spring constant given by $k_{eff} = \kappa/L_{Sph}^2$ These results are summarized in Table 1 above.

## APPENDIX D: Minimum separation due to Casimir pull-in

The minimum stable separation for a linear spring plate-sphere setup has been derived in the main text. Here, the derivation is given for a linear spring parallel-plate setup followed by a torsional spring plate-sphere setup.

For two parallel plates separated by a spring and in the presence of the Casimir force, the generalized spring constant becomes:

$$k_{sys} = -\frac{\partial F_{tot}}{\partial z} = \frac{\pi^2 \hbar c}{60} \frac{A}{(z_0-z)^5} - k_0 = 0. \quad (D1)$$

As the spring force is balanced by the Casimir force: $k_0 = \frac{\pi^2 \hbar c}{240} \frac{zA}{(z_0-z)^3}$. Hence, one finds that the pull-in occurs at $z = \frac{z_0}{5}$, and the minimal stable plate separation results in:

$$z_{0C-PP} = \sqrt[5]{\frac{625\,\pi^2 \hbar c}{12288}\frac{A}{k_0}} = \gamma_{CP-PP}\sqrt[5]{\frac{A}{k_0}}, \tag{D2}$$

where now $\gamma_{CP-PP} = \sqrt[5]{\frac{625\,\pi^2 \hbar c}{12288}} \cong 6.848 \times 10^{-6}$ N$^{\frac{1}{5}}$m$^{\frac{2}{5}}$ is the pull-in constant for the Casimir-spring double-plate system. (This can be generalized to find a $z_0/n$ pull-in for linear restoring forces that balance $1/(z_0\text{-}z)^{n\text{-}1}$ attractive forces.) Using $A = 100 \times 100$ μm$^2$, $k_0 = 0.2$ N/m results in $z_{0c-PP} = 240$ nm. This sets a hard limit on the parameter space of stable operation of the parametric amplifier.

The argument for the torsional pull-in position is very similar to that given for the linear pull-in. The total torque is zero for static equilibrium at all distances greater than the pull-in. For small angles the vertical distance can be approximated as $z_0 - L_{Sph}\theta$ leading to:

$$\tau_{tot} = 0 = -\frac{\pi^3 \hbar c\, R}{360\,(z_0 - L_{Sph}\theta)^3}L_{Sph} + \kappa_0 \theta, \tag{D3}$$

where $\kappa_0$ is the torsional spring constant. The torsional spring constant of the entire system can be written as:

$$\kappa_{sys} = -\frac{\partial \tau_{tot}}{\partial \theta} = \frac{\pi^3 \hbar c\, R}{120\,(z_0 - L_{Sph}\theta)^4}L_{Sph}^2 - \kappa_0. \tag{D4}$$

Again, the closest stable approach is at $\kappa_{sys} = 0$. We can then solve for the spring constant $\kappa_0$ from $\kappa_{sys}$ and evaluate the static torque $\tau_{tot} = 0$ with this value of $\kappa_0$. The pull-in angle is then given as:

$$\theta = \frac{z_0}{4\,L_{Sph}}. \tag{D5}$$

This yields the same result as the vertical pull-in and the small angle approximation. Substituting this angle into $\tau_{tot}$ for $\theta$, and solving for $z_0$ we find the pull-in constant $\gamma_{CP}$ (from Equation 6) for the Casimir-spring, plate-sphere system remains the same. However, the position now depends on $R, L_{Sph}$ and $\kappa_0$ and is given as:

$$z_{0C-PS}(Torsional) = \sqrt[4]{\frac{32\,\pi^3 \hbar c}{1215}\frac{R L_{Sph}^2}{\kappa_0}}$$

$$= \gamma_{CP}\sqrt[4]{\frac{R\,L_{Sph}^2}{\kappa_0}} \tag{D6}$$

Thus with a torsional plate one can tune the pull-in position by moving the sphere closer to or farther from the pivot point $L_{Sph}$.

### APPENDIX E: Thermal Noise in a torsion resonator

The minimum theoretical value for $z_0$ is dictated by the Casimir pull-in separation described above. For large gain, the amplitude $z_a$ must be limited so that the condition $z_a \ll z_0$ still holds (otherwise higher order terms in the derivation of the mechanical response need to be considered). Naturally $z_a$ must always be greater than the mechanical thermal noise amplitude $z_{th}$, resulting in a condition for the minimum possible $z_a$ and corresponding maximum allowed gain.

The thermal-mechanical spectral density of the angular amplitude for torsion resonator is given by:

$$S_\theta^{1/2}(\omega) = \sqrt{\frac{4 k_B T \omega_0}{Q}\frac{1}{I\left[(\omega^2-\omega_0^2)^2+\left(\frac{\omega \omega_0}{Q}\right)^2\right]}}, \tag{E1}$$

where $k_B$ is the Boltzmann constant, $T$ the temperature, and $Q$, $\omega_0$ and $I$ are the quality factor, resonance frequency and moment of inertia of the torsional resonator respectively. The Casimir amplifier described above will be tuned on resonance. Given a rectangular rod the on resonance spectral density of the angular amplitude becomes [55]:

$$S_\theta^{1/2} = \sqrt[4]{\frac{k_B^2 T^2 Q^2 \rho}{\beta^3 G_{Si}^3} \frac{L_p L_s^3 w_p^3}{t_s^2 w_s^9}}, \tag{E2}$$

where $k_B$ is the Boltzmann constant, $T$ the temperature, $\rho = 2320$ kgm$^{-3}$ the density of polysilicon, $\beta = 2.43$ for $w_s/t_s = 1.5$ a numeric constant characterizing the torsion spring, $G_{Si} = 69$ GPa the shear modulus of poly-silicon, $t_s = 2$ µm is the spring thickness, and $L_S = 40$ µm and $w_S = 3$ µm the length and width respectively of the spring (*s*) and plate (*p*). The relevant spectral density displacement amplitude is a result of the geometric position of the metal sphere, $L_{sph} = 175$ µm, from the torsion axis such that on resonance:

$$S_{z_{th}}^{1/2}(\omega_0) = L_{sph} \sin\left(\sqrt[4]{\frac{k_B^2 T^2 Q^2 \rho}{\beta^3 G_{Si}^3} \frac{L_p L_s^3 w_p^3}{t^2 w_s^9}}\right) \approx L_{sph} \sqrt[4]{\frac{k_B^2 T^2 Q^2 \rho}{\beta^3 G_{Si}^3} \frac{L_p L_s^3 w_p^3}{t^2 w_s^9}}, \tag{E3}$$

where the linear approximation of the sine function is valid only for small angles. Higher-order mechanical modes exist, but these are suppressed as $\omega_0^{-3/2}$. It is assumed here that the torsion mode dominates and, for the resonant response, only eqn. (E2) must be considered. From the Equipartition Theorem, one can predict a root-mean-square displacement on the order of:

$$\Delta z_{th} = \sqrt{\frac{k_B T}{m \omega_0^2}}. \tag{E4}$$

This expression sets the bandwidth-independent expected amplitude of a finite temperature resonator. It can serve as a minimum-allowed resonator amplitude. At $T = 300$ K the thermal amplitude is $\Delta z_{th} \cong 19$ pm for the plate resonator considered here, and at $T = 4.2$ K the thermal amplitude drops to $\Delta z_{th} \cong 2$ pm. For $z_0 = 100$ nm, this results in a maximum gain of ~500 and 5000 respectively. Multiplying eqn. (E3) by the square root of the bandwidth of the resonator ($BW \sim \frac{\omega_0}{Q}$) increases the thermal motion compared to eqn. (E4) by a factor ~2.